\documentclass[letterpaper, 10 pt, onecolumn, conference, draftcls]{ieeeconf}  

\IEEEoverridecommandlockouts                              
\usepackage{graphicx}
\usepackage{epsfig} 
\usepackage{fancyhdr}
\usepackage{stfloats}
\usepackage{amsmath} 
\usepackage{amssymb}  
\usepackage{mathrsfs}
\usepackage{url}
\usepackage{tabularx,ragged2e,booktabs,caption}
\usepackage{bm}
\usepackage{bbm}
\usepackage{color}
\usepackage{theorem}
\usepackage{caption}
\usepackage{subcaption}
\usepackage{supertabular}
\usepackage{hyperref}
\usepackage{float}
\usepackage{longtable}
\usepackage{array} 
\allowdisplaybreaks

\usepackage{algorithm}
\usepackage{algorithmicx,algpseudocode}

\newcommand{\be}[1]{\begin{equation}\label{#1}}
\newcommand{\benon}{\begin{equation*}}  
\newcommand{\bemuln}[1]{\begin{multline}\label{#1}}
\newcommand{\bemul}{\begin{multline*}}
\newcommand{\bee}{\begin{eqnarray*}}
\newcommand{\eee}{\end{eqnarray*}}
\newcommand{\been}[1]{\begin{eqnarray}\label{#1}}
\newcommand{\eeen}{\end{eqnarray}}
\newcommand{\began}[1]{\begin{gather}\label{#1}}
\newcommand{\bega}{\begin{gather*}}
\newcommand{\bealn}[1]{\begin{align}\label{#1}}
\newcommand{\beal}{\begin{align*}}
\newcommand{\bealatn}[2]{\begin{alignat}{#1}\label{#2}}
\newcommand{\bealat}{\begin{alignat*}}
\newcommand{\bexalatn}[1]{\begin{xalignat}\label{#1}}
\newcommand{\bexalat}{\begin{xalignat*}}

\newcommand{\mb}{\mathbf}

{\theoremstyle{plain} \newtheorem{thm}{Theorem}[section]

\newtheorem{rem}{Remark}

}
{\theoremstyle{break} \theorembodyfont{\it}
\newtheorem{defi}{Definition}
\newtheorem{ass}{Assumption} 
 }

\def\bd{{\mathbf d}}

\def\bg{{\mathbf g}}

\def\bt{{\mathbf t}}

\def\bw{{\mathbf w}}
\def\bx{{\mathbf x}}  
\def\by{{\mathbf y}}

\def\bN{{\mathbf N}}

\def\texitem#1{\par\smallskip\noindent\hangindent 25pt
               \hbox to 25pt {\hss #1 ~}\ignorespaces}

\newcommand{\scrA}{\mathcal{A}}

\newcommand{\scrD}{\mathcal{D}}

\newcommand{\scrF}{\mathcal{F}}

\newcommand{\scrH}{\mathcal{H}}

\newcommand{\scrK}{\mathcal{K}}

\newcommand{\scrU}{\mathcal{U}}
\newcommand{\scrV}{\mathcal{V}}
\newcommand{\scrW}{\mathcal{W}}

\newcommand{\bbeta}{\boldsymbol{\beta}}

\newcommand{\bepsilon}{\boldsymbol{\epsilon}}

\newcommand{\btheta}{\boldsymbol{\theta}}

\DeclareMathOperator{\VI}{\text{VI}}

\makeatletter

\newcommand{\scrRmnum}[1]{\expandafter\@slowromancap\romannumeral #1@}
\newcommand*{\defeq}{\stackrel{\text{def}}{=}}
\makeatother

\title{\LARGE \bf Data-Driven Estimation of Travel Latency Cost Functions via Inverse Optimization in Multi-Class Transportation Networks\authorrefmark{1} 
	\thanks{* Research partially supported by the NSF under grants CCF-1527292
		and IIS-1237022, by the ARO under grant W911NF-12-1-0390 and by the
		Cyprus Research Promotion Foundation under Grant New Infrastructure
		Project/Strategic/0308/26.}
}

\author{Jing Zhang$^\dag$ and Ioannis Ch. Paschalidis$^\ddag$
	\thanks{$\dag$ Division of Systems Eng., Boston University, e-mail:
		{\tt jzh@bu.edu}.}
	\thanks{$\ddag$ Department of Electrical and
		Computer Eng., Division of Systems Eng., and Dept. of Biomedical Eng., Boston University,
		8 St. Mary's St., Boston, MA 02215, e-mail: {\tt yannisp@bu.edu, http://sites.bu.edu/paschalidis/}.}}

\begin{document}

	\maketitle

	\begin{abstract}
		We develop a method to estimate from data travel
		latency cost functions in multi-class transportation networks,
		which accommodate different types of vehicles with very different characteristics (e.g., cars and trucks). Leveraging our earlier
		work on inverse variational inequalities, we develop a data-driven approach to estimate the travel latency cost functions.
		Extensive numerical experiments using benchmark networks,
		ranging from moderate-sized to large-sized, demonstrate the
		effectiveness and efficiency of our approach.
		
	\end{abstract}
	

	\section{Introduction}

	Assuming all users (drivers) in a transportation network select routes selfishly, the network ends up reaching an equilibrium in terms of link flows, which is known as the Wardrop Equilibrium (WE) \cite{wardrop1952some}, an instantiation of the well-known Nash Equilibrium. At the equilibrium, no single driver can ``benefit'' by rerouting. 
	Mathematically, given \emph{travel latency (time) cost functions} (we will also simply say \emph{cost functions} in the sequel) with respect to link (road/arc) flows and an origin-destination (OD) flow demand matrix, finding the WE is formulated as the Traffic Assignment Problem (TAP), which has been well-studied; see, e.g., \cite{patriksson1994traffic} and the references therein. We will call TAP the \emph{forward problem} throughout the paper. On the other hand, given an OD demand matrix and the WE, one can seek to ``recover'' the cost functions, by formulating appropriate \emph{inverse problems}; see, e.g., \cite{bertsimas2014data,CDC16,IFAC17}, among others. Moreover, another version of the inverse problem can be considered: estimating/adjusting the OD demand matrix by assuming that the WE and the cost functions are known (sometimes a ``rough'' initial OD demand matrix is also available); see, e.g., \cite{spiess1990gradient,lundgren2008heuristic}. In this paper, applying a data-driven approach, we only consider the former version \emph{inverse problems}, namely, recovering the cost functions from a given OD demand matrix and the corresponding WE, which is of practical importance, as partly shown in \cite{CDC16,IFAC17}. 
	
	Depending on whether we put different weights onto the flows from different types of vehicles, we can model a transportation network as \emph{single-class} \cite{dafermos1969traffic} or \emph{multi-class} \cite{dafermos1972traffic}. The \emph{forward problem} (i.e., the TAP) for both single-class and multi-class transportation networks has been widely studied; see, e.g., \cite{dafermos1969traffic, dafermos1972traffic}, among others. In addition, recently the \emph{inverse problem} for the single-class transportation networks has been investigated as well \cite{bertsimas2014data,CDC16,IFAC17}. However, to the best knowledge of the authors, there has been little work on the \emph{inverse problem} for multi-class transportation networks. Our goal in this paper, then, is to fill this gap.
	
	It is a well-known fact that, for multi-class transportation networks, there do not exist reasonable easily verifiable assumptions about the cost functions  to ensure the existence and uniqueness of the solution to the \emph{forward problem} \cite{noriega2007algorithmic}. Therefore, for the multi-class \emph{inverse problem}, it would be hard to establish rigorous theoretical results under mild easy-to-check conditions. We, in turn, seek to first propose an appropriate formulation for the multi-class \emph{inverse problem}, and then empirically validate our solution by conducting extensive numerical experiments.
	
	The rest of the paper is organized as follows. We present the transportation network model in Sec. \ref{sec:mod}. In Sec. \ref{sec:prob} we formulate the multi-class \emph{forward problem} (TAP), specify the form of the cost functions, and formulate the multi-class \emph{inverse problem}. Numerical results are shown in Sec. \ref{sec:num}. Sec. \ref{sec:con} concludes the paper.
	
	\textbf{Notational conventions:} All vectors are column vectors. For
	economy of space, we write $\mathbf x = (x_1, \ldots, x_{\text{dim}(\bx)})$ to denote
	the column vector $\bx$, where $\text{dim}(\bx)$ is its dimensionality. $\bx \ge \textbf{0}$ (with $\textbf{0}$ being the zero vector) denotes that all entries of a vector $\bx$ are nonnegative. Denote by $\mathbb{R}_+$ the set of all nonnegative real numbers. We use ``prime'' to denote the transpose of a
	matrix or vector. Unless otherwise specified, $\|\cdot\|$ denotes the
	$\ell_2$ norm. Let $\left| \mathcal{D} \right|$ denote the cardinality of a
	set $\mathcal{D}$, and $\left[\kern-0.15em\left[ \scrD 
	\right]\kern-0.15em\right]$ the set $\left\{ {1, \ldots ,\left| \scrD \right|} \right\}$. $A \defeq B$ indicates $A$ is defined using
	$B$.
	
	\section{Transportation Network Model}  \label{sec:mod}
	
	\subsection{Single-class transportation network model} \label{sec:sing-mod}
	
	Consider a directed graph, denoted by $\left( {\scrV, \scrA}
	\right)$, where $\scrV$ denotes the set of nodes and $\scrA$ the
	set of links. Assume it is strongly connected.  Let $\textbf{N} \in {\left\{
		{0,1, - 1} \right\}^{\left| \scrV \right| \times \left| \scrA
			\right|}}$ be the node-link incidence matrix, and $\textbf{e}_{a}$ the
	vector with an entry equal to 1 corresponding to link $a$ and all the other
	entries equal to 0.
	
	Let $\bw = \left( {{w_s},{w_t}} \right)$ denote an origin-destination
	(OD) pair and $\scrW = \left\{ {{\bw_i}:{\bw_i} = \left(
		{{w_{si}},{w_{ti}}} \right), \,i \in \left[\kern-0.15em\left[ \scrW 
		\right]\kern-0.15em\right]}
	\right\}$ the set of all OD pairs.  Denote by ${d^{\bw}} \ge 0$ the
	amount of the flow demand from $w_s$ to $w_t$. Let ${\bd^{\bw}}
	\in {\mathbb{R}^{\left| \scrV \right|}}$ be the vector which is all
	zeros, except for a $-{d^{\bw}}$ in the coordinate corresponding to node
	$w_s$ and a ${d^{\bw}}$ in the coordinate corresponding to node $w_t$.
	
	
	Let $x_a$ be the total link flow on link $a \in \scrA$ and $\bx$
	the vector of these flows.
	Let $\scrF$ be the set of feasible flow vectors defined by
	\begin{align}
	\begin{array}{l}
	\scrF \defeq \Big\{ {\bx:\exists {\bx^{\bw}} \in \mathbb{R}_ +
		^{\left| \scrA \right|} ~\text{s.t.}~\bx =
		\sum\limits_{\bw \in \scrW} {{\bx^{\bw}}},} \Big. \label{feasi} \\
	~~~~~~~~~\Big. \textbf{N}{\bx^{\bw}} = {\bd^{\bw}},\,\forall \bw \in \scrW \Big\}, 
	\end{array}  
	\end{align}
	where $\bx^\bw$ is the flow vector attributed to OD pair $\bw$.
	

	\subsection{Multi-class transportation network model} \label{sec:multi-mod}
	
	Denote by $|\tilde \scrU|$ the number of user (vehicle) classes. Let the original network be $\big( {\tilde {\scrV},\tilde {\scrA}}, \tilde {\scrW} \big)$,  where 
	$\tilde {\scrV} = \big\{ {{v_1}, \ldots ,{v_{|\tilde \scrV|}}} \big\}$,
	$\tilde {\scrA} = \big\{ {{a_{1}}, \ldots ,{a_{|\tilde \scrA|}}} \big\}$, and $\tilde {\scrW} = \big\{ {{\bw_i}:{\bw_i} = \left( {{w_{si}},{w_{ti}}} \right),i \in [\kern-0.15em[ \tilde \scrW ]\kern-0.15em]} \big\}$.
	We borrow the idea of making $|\tilde \scrU|$ copies of $\big( {\tilde {\scrV},\tilde {\scrA}}, \tilde {\scrW} \big)$, each corresponding to a single user class, to obtain an enlarged single-class network \cite{dafermos1972traffic}. In particular, we construct a single-class network $\left( \scrV, \scrA,\scrW \right)$, where
	\begin{align}
	\scrV &= \big\{ {v(i,u): i \in [\kern-0.15em[ \tilde\scrV 
		]\kern-0.15em], u \in [\kern-0.15em[ \tilde\scrU 
		]\kern-0.15em]} \big\}, \notag \\
	\scrA &= \big\{ {a(i,u): i \in [\kern-0.15em[ \tilde\scrA 
		]\kern-0.15em], u \in [\kern-0.15em[ \tilde\scrU 
		]\kern-0.15em]} \big\}, \notag \\
	\scrW &= \big\{ \bw(i,u):\bw(i,u) = \left( {w_s(i,u),w_t(i,u)} \right), \big. \notag 
	\\ &~~~~~~\big. i \in [\kern-0.15em[ \tilde\scrW 
	]\kern-0.15em], \,u \in [\kern-0.15em[ \tilde\scrU 
	]\kern-0.15em] \big\}. \notag
	\end{align}

	We then can write the node-link incidence matrix $\textbf{N} \in {\left\{ {0,1, - 1} \right\}^{\left| \scrV \right| \times \left| \scrA \right|}}$. Note that $\left| \scrV \right| = |\tilde\scrU||\tilde\scrV|$, and $\left| \scrA \right| = |\tilde\scrU||\tilde\scrA|$. For $i \in [\kern-0.15em[ \tilde\scrA 
	]\kern-0.15em],~ u \in [\kern-0.15em[ \tilde\scrU 
	]\kern-0.15em]$, let $\textbf{e}_{iu}$ denote the $\left| \scrA \right|$-dimensional
	vector with an entry equal to 1 corresponding to link $a(i,u)$ and all the other
	entries equal to 0.
	Also, for $i \in [\kern-0.15em[ \tilde\scrW 
	]\kern-0.15em],~ u \in [\kern-0.15em[ \tilde\scrU 
	]\kern-0.15em]$ we denote by ${d^{\bw(i,u)}}$ the flow demand of user class $u$ from node
	${w_s(i,u)}$ to node ${w_t(i,u)}$. Then, we define ${\bd^{\bw}} \in {\mathbb{R}^{\left| \scrV \right|}}$ as for the single-class case.
	Accordingly, the set of feasible flow vectors $\scrF$ can be defined as the one in \eqref{feasi}.
	For convenience, we also denote by $\bg^{{(u)}} \in {\mathbb{R}^{| \tilde\scrW |}}$ the vectorized OD demand matrix for user class $u \in [\kern-0.15em[ \tilde\scrU 
	]\kern-0.15em]$.

	Writing a feasible flow vector $\bx \in \scrF$ as
	$\bx = \big( {x_{iu}; \,i \in [\kern-0.15em[ \tilde\scrA 
		]\kern-0.15em], u \in [\kern-0.15em[ \tilde\scrU 
		]\kern-0.15em]} \big)$, where $x_{iu}$ denotes the flow on link $a(i,u)$,
	we consider the following cost function (the cost on a \emph{physical} link does not depend on the flows elsewhere; a \emph{physical} link maps to $|\tilde{\scrU}|$ \emph{conceptual} links, each of which corresponds to a user class):
	\begin{align}
	\bt\left( \bx \right) = \big({t_{iu}}( {{x_{i1}}, \ldots ,{x_{i|\tilde\scrU|}}} ); \,i \in [\kern-0.15em[ \tilde\scrA 
	]\kern-0.15em], u \in [\kern-0.15em[ \tilde\scrU 
	]\kern-0.15em]\big). \label{costf} 
	\end{align}
	
	\section{Problem Formulation} \label{sec:prob}
	
	\subsection{The multi-class forward VI problem}  \label{sec:forward}

	As in \cite{CDC16}, here we refer to the Traffic Assignment Problem (TAP) as the \textit{forward problem}, whose goal is to find the
	Wardrop Equilibrium for a given single-class transportation network with a given travel latency cost function and a
	given OD demand matrix.  
	It is a well-known result that the TAP can be formulated as a Variational Inequality (VI) problem $\text{VI}\left( {\bt,\scrF} \right)$, defined as follows:
	
	\begin{defi}
		\label{cdc16-def2} \em{(\cite{bertsimas2014data})}. The VI problem, denoted as $\text{VI}\left( {\bt,\scrF} \right)$, is to find an ${\bx^ * } \in \scrF$ s.t.
		\begin{align}
		\bt{\left( {{\bx^ * }} \right)'}\left( {\bx - {\bx^ * }} \right) \geq 0, \quad \forall \bx \in \scrF. \label{cdc16-VI}
		\end{align}
	\end{defi}
	
	Let us present a definition regarding the monotonicity of a cost function:
	
	\begin{defi}
		\label{journal-def3} \em{(\cite{patriksson1994traffic})}. 
		$\bt(\cdot)$ is \textit{strongly monotone} on $\scrF$ if there exists a constant $\eta > 0$ such that
		\[\left( {\bt\left( \bx \right) - \bt\left( \by \right)} \right)'\left( {\bx - \by} \right) \ge \eta {\left\| {\bx - \by} \right\|^2},\quad \forall \bx,\by \in \scrF.\]
	\end{defi}
	To ensure the existence and uniqueness of the solution to $\text{VI}\left( {\bt,\scrF} \right)$, we need the following assumption:
	\begin{ass}
		\label{cdc16-assumption1}
		$\bt(\cdot)$ is strongly monotone on $\scrF$ and
		continuously differentiable on $\mathbb{R}_ + ^{\left| \scrA
			\right|}$. $\scrF$ is nonempty and contains an interior point
		(Slater's condition).
	\end{ass}

	For the enlarged network $\left( \scrV, \scrA,\scrW \right)$, the VI result for the single-class transportation network applies. In particular, we have:
	
	\begin{thm} \label{cdc17-th21} (\cite{patriksson1994traffic}).
		\noindent Under Assump. \ref{cdc16-assumption1}, a Wardrop Equilibrium of the multi-class transportation network is a solution to $\text{VI}(\bt,\scrF)$, where $\bt,\scrF$ are given by \eqref{costf} and \eqref{feasi}, respectively.
	\end{thm}

	\begin{rem} \label{rem1} \em{As noted in \cite{noriega2007algorithmic}, Assump. A cannot be easily verified for general multi-class transportation networks. We therefore do not have any guarantee of always obtaining unique link flows for each and every type of users (vehicles). Thus, the following results turn out to be empirical. However, as we will note in Remark \ref{rem2}, we still have great hope to recover the cost functions with good accuracy from the weighted sum of link flows of different types of users.}
		
	\end{rem}

	\subsection{BPR-type cost functions for multi-class transportation network} \label{sec:cost}
	
	We now further specify the cost functions in \eqref{costf}.
	For each $i \in [\kern-0.15em[ \tilde\scrA 
	]\kern-0.15em], u \in [\kern-0.15em[ \tilde\scrU 
	]\kern-0.15em]$, we define the following generalized Bureau of Public Roads (BPR)-type travel time function \cite{branston1976link, noriega2007algorithmic, bertsimas2014data}:
	\begin{align}{t_{iu}}\left( \bx \right) = {t^{{0}}_{iu}}f\left( {\frac{{{\btheta}'{{\bx_{i}}} }}{{{m_{i}}}}} \right), \label{costMulti}
	\end{align}
	where ${t^{{0}}_{iu}}$ is called the \textit{free-flow travel time} of link ${a(i,u)}$, $f(0)=1$, $f(\cdot)$ is strictly increasing and continuously
	differentiable on $\mathbb{R}_+$, ${m_{i}}$  is the \textit{effective flow capacity} of the $i$th \emph{physical} link, ${\bx_{i}} = \big( {x_{iu}};\, u \in [\kern-0.15em[ \tilde\scrU 
	]\kern-0.15em]\big)$ is the flow vector of all user classes corresponding to the $i$th \emph{physical} link, and ${\btheta} = \big( {\theta _u}; \, u \in [\kern-0.15em[ \tilde\scrU 
	]\kern-0.15em] \big)$ is a weight vector such that ${\theta _u} \geq 0, \,\forall u \in [\kern-0.15em[ \tilde\scrU 
	]\kern-0.15em]$.

	\subsection{The multi-class inverse VI problem formulation} \label{sec InverseVI-multi}
	
	To solve the forward problem, we need to know the cost function and the OD demand matrix. Assuming that we know the OD demand matrix and have observed the Wardrop Equilibrium flows, we seek to formulate the inverse problem (the inverse VI problem, in particular), so as to estimate the travel latency cost function (specifically, $f(\cdot)$).
	
	For a given $\epsilon > 0$, we define an
	\textit{$\epsilon$-approximate solution} to $\VI(\mathbf{t}, \scrF)$ by
	changing the right-hand side of \eqref{cdc16-VI} to $- \epsilon$:
	
	\begin{defi}
		\label{cdc16-def3}
		\em{(\cite{bertsimas2014data})}. Given $\epsilon > 0$,
		$\hat{\bx} \in \scrF$ is called an
		\textit{$\epsilon$-approximate solution} to $\VI(\mathbf{t}, \scrF)$ if
		\begin{equation}
		\label{cdc16-eq:defApproxEquil}
		\mathbf{t}(\hat{\bx})'(\bx - \hat{\bx}) \geq - \epsilon, \quad
		\forall \bx \in \scrF.
		\end{equation}
	\end{defi}

	Given $|\scrK|$ observations $(\bx^{{(k)}}, \scrF^{{(k)}})$, $k \in [\kern-0.15em[ \scrK 
	]\kern-0.15em]$,
	with $\bx^{{(k)}} \in \scrF^{{(k)}}$ and each $\scrF^{{(k)}}$ being a set of feasible flow
	vectors meeting Slater's condition \cite{boyd2004convex}, the inverse VI problem
	amounts to finding a function $\bt$ such that $\bx^{{(k)}}$ is an
	$\epsilon_k$-approximate solution to $\VI(\bt, \scrF^{{(k)}})$ for each $k$.
	Denoting $\bepsilon \defeq (\epsilon_k; \, k \in [\kern-0.15em[ \scrK 
	]\kern-0.15em])$, we can formulate the inverse VI problem as \cite{bertsimas2014data}
	\begin{align}
	\min_{{\mathbf{t}}, \bepsilon} \quad & \| \bepsilon \|
	\label{cdc16-inverseVI1} \\
	\text{s.t.} \quad & {\mb{t}}({\bx^{{(k)}}})'(\bx-{\bx^{{(k)}}}) \geq -\epsilon_k, \quad \forall
	\bx\in \scrF^{{(k)}}, k \in [\kern-0.15em[ \scrK 
	]\kern-0.15em], \notag \\
	& \epsilon_k > 0, \quad \forall k \in [\kern-0.15em[ \scrK 
	]\kern-0.15em]. \notag
	\end{align}

	\begin{figure*}[ht]
		\begin{align}
		\text{(invVI-1)} \quad
		\quad \mathop {\min }\limits_{\by,\bepsilon} {\text{  }}& \|\bepsilon\| + \gamma\| {f} \|^2_{\scrH} \label{inverVI2-multi-pre} \\
		\text{s.t.}{\text{  }}&{\textbf{e}}_{iu}'\bN'_k{\by^\bw} \leq t_{iu}^{{0}}{{{f}\left( \frac{{{{\btheta}}'\bx_{{i}}^{{(k)}}}}{{{m_{i}^{{(k)}}}}} \right)}}, \qquad \forall i \in [\kern-0.15em[ \tilde\scrA^{{(k)}} 
		]\kern-0.15em], ~u \in [\kern-0.15em[ \tilde\scrU 
		]\kern-0.15em], ~\bw \in {\scrW^{{(k)}}}, ~k \in [\kern-0.15em[ \scrK 
		]\kern-0.15em],  \label{dual-feasi} \\
		&\sum\limits_{i=1}^{|\tilde\scrA^{{(k)}}|} \Bigg(\sum\limits_{u=1}^{|\tilde\scrU|} {t_{iu}^{{0}}{x_{iu}}}\Bigg) {{{f}\left( \frac{{{{\btheta}}'\bx_{{i}}^{{(k)}}}}{{{m_{i}^{{(k)}}}}} \right)}} - \sum\limits_{\bw \in {\scrW_k}} {\left( {{\bd^\bw}} \right)'{\by^\bw}}  \leq {\epsilon_k},  \qquad \forall k \in [\kern-0.15em[ \scrK 
		]\kern-0.15em],  \label{sub-opt} \\
		&{ {{f}\left( \frac{{{{\btheta}}'\bx_{{i}}^{{(k)}}}}{{{m_{i}^{{(k)}}}}} \right)}} \leq {{{f}\bigg( \frac{{{{\btheta}}'\bx_{{{\tilde i}}}^{{(k)}}}}{{{m^{{(k)}}_{{\tilde i}}}}} \bigg)}}, \qquad \forall i, \, \tilde{i} \in [\kern-0.15em[ \tilde\scrA^{{(k)}} 
		]\kern-0.15em]
		~{\text{ s}}{\text{.t}}{\text{. }}\frac{{{{\btheta}}'\bx_{{i}}^{{(k)}}}}{{{m_{i}^{{(k)}}}}} \leq \frac{{{{\btheta}}'\bx_{{{\tilde i}}}^{{(k)}}}}{{{m^{{(k)}}_{{\tilde i}}}}}; \,\forall k \in [\kern-0.15em[ \scrK 
		]\kern-0.15em],  \label{monoto} \\
		& \bepsilon \geq \textbf{0}, \quad 
		{f} \in \scrH, \notag \\
		&{f(0)} = 1. \label{normaliz} 
		\end{align}
		\hrulefill
	\end{figure*}
	
	\begin{figure*}[ht]
		\begin{align}
		\text{(invVI-2)} \quad
		\quad \mathop {\min }\limits_{\bbeta ,\by,\bepsilon} {\text{  }}& \|\bepsilon\| + \gamma \sum\limits_{j = 0}^n {\frac{{\beta _j^2}}{{{n \choose j}{c^{n - j}}}}} \label{inverVI2-multi} \\
		\text{s.t.}{\text{  }}&{\textbf{e}}_{iu}'\bN'_k{\by^\bw} \leq t_{iu}^{{0}}{\sum\limits_{j = 0}^n {{\beta _j}\left( \frac{{{{\btheta}}'\bx_{{i}}^{{(k)}}}}{{{m_{i}^{{(k)}}}}} \right)} ^j}, \qquad \forall i \in [\kern-0.15em[ \tilde\scrA^{{(k)}} 
		]\kern-0.15em], ~u \in [\kern-0.15em[ \tilde\scrU 
		]\kern-0.15em], ~\bw \in {\scrW^{{(k)}}}, ~k \in [\kern-0.15em[ \scrK 
		]\kern-0.15em],  \notag \\
		&\sum\limits_{i=1}^{|\tilde\scrA^{{(k)}}|} {\Bigg({\sum\limits_{j = 0}^n {{\beta _j}\bigg( \frac{{{{\btheta}}'\bx_{{i}}^{{(k)}}}}{{{m_{i}^{{(k)}}}}} \bigg)} ^j}\Bigg)} \sum\limits_{u=1}^{|\tilde\scrU|} {t_{iu}^{{0}}{x_{iu}}} - \sum\limits_{\bw \in {\scrW_k}} {\left( {{\bd^\bw}} \right)'{\by^\bw}}  \leq {\epsilon_k},  \qquad \forall k \in [\kern-0.15em[ \scrK 
		]\kern-0.15em],  \notag \\
		&{\sum\limits_{j = 0}^n {{\beta _j}\bigg( \frac{{{{\btheta}}'\bx_{{i}}^{{(k)}}}}{{{m_{i}^{{(k)}}}}} \bigg)} ^j} \leq {\sum\limits_{j = 0}^n {{\beta _j}\bigg( \frac{{{{\btheta}}'\bx_{{{\tilde i}}}^{{(k)}}}}{{{m^{{(k)}}_{{\tilde i}}}}} \bigg)} ^j}, \qquad \forall i, \, \tilde{i} \in [\kern-0.15em[ \tilde\scrA^{{(k)}} 
		]\kern-0.15em]
		~{\text{ s}}{\text{.t}}{\text{. }}\frac{{{{\btheta}}'\bx_{{i}}^{{(k)}}}}{{{m_{i}^{{(k)}}}}} \leq \frac{{{{\btheta}}'\bx_{{{\tilde i}}}^{{(k)}}}}{{{m^{{(k)}}_{{\tilde i}}}}}; \,\forall k \in [\kern-0.15em[ \scrK 
		]\kern-0.15em],  \notag \\
		& \bepsilon \geq \textbf{0}, \notag \\
		&{\beta _0} = 1. \notag 
		\end{align}
		\hrulefill
	\end{figure*}

	Assume now we are given $|\scrK|$ networks $(\scrV^{{(k)}},\scrA^{{(k)}},\scrW^{{(k)}}),\, k \in [\kern-0.15em[ \scrK 
	]\kern-0.15em]$ (as a special case, these could be $|\scrK|$ replicas of the same network $\left( {\scrV, \scrA, \scrW} \right)$), and the observed link flow data $\big( {\bx_{{i}}^{{(k)}}} = \big( {x^{{(k)}}_{iu}};\, u \in [\kern-0.15em[ \tilde\scrU 
	]\kern-0.15em]\big); \, i \in [\kern-0.15em[ \tilde\scrA^{{(k)}} 
	]\kern-0.15em], k \in [\kern-0.15em[ \scrK 
	]\kern-0.15em]\big)$. Aiming at recovering a cost function that has both good data reconciling and generalization properties, to make \eqref{cdc16-inverseVI1} solvable, we apply an estimation approach
	which expresses the function $f(\cdot)$ (recall \eqref{costMulti}) in a Reproducing Kernel Hilbert Space
	(RKHS) $\scrH$ \cite{bertsimas2014data,evgeniou2000regularization}. In particular, by \cite[Thm. 2]{bertsimas2014data}, being a variant of \cite[(6)]{IFAC17}, the inverse VI problem \eqref{cdc16-inverseVI1} can be reformulated as a Quadratic Program (QP) invVI-1 (see \eqref{inverVI2-multi-pre}), where $\by = \big( {{\by^{\bw} \in {\mathbb{R}^{| \scrV^{{(k)}} |}}};\, \bw \in \scrW^{{(k)}}, k \in [\kern-0.15em[ \scrK 
		]\kern-0.15em]} \big)$, and $\bepsilon = (\epsilon_k; \,k \in [\kern-0.15em[ \scrK 
	]\kern-0.15em])$ are decision vectors, $\gamma > 0$ is a regularization parameter (a smaller $\gamma$ should result in recovering a ``tighter'' $f(\cdot)$ in terms of data reconciling; a bigger $\gamma$, on the other hand, would lead to a ``better'' $f(\cdot)$ in terms of generalization properties), $\| {f} \|^2_{\scrH}$ denotes the squared norm of $f(\cdot)$ in $\scrH$, \eqref{dual-feasi} is for dual feasibility, \eqref{sub-opt} is the suboptimality (primal-dual gap) constraint, \eqref{monoto} enforces $f(\cdot)$ to be nondecreasing, and \eqref{normaliz} is for normalization purposes (see \eqref{costMulti}).
	
	It can be seen that the above formulation is still too abstract for us to solve, because it is an optimization over functions. To make it tractable, in the following, we will specify $\scrH$ by picking its \textit{reproducing kernel} \cite{evgeniou2000regularization} as a polynomial ${\phi}(x, y) \defeq ( c + xy )^n$ for some choice of
	$c \geq 0$ and $n \in \mathbb{N}$. Then, writing 
	\[\phi\left( {x,y} \right) = {\left( {c + xy} \right)^n} = \sum\limits_{j = 0}^n {{n \choose j}{c^{n - j}}{x^j}{y^j}}, \]
	by \cite[(3.2), (3.3), and (3.6)]{evgeniou2000regularization},
	we instantiate invVI-1 as a QP invVI-2 (see \eqref{inverVI2-multi}),
	which involves $\bbeta  = \left( {{\beta _j};\,j = 0, 1, \ldots ,n} \right)$ as an additional decision vector.
	Assuming an optimal ${\bbeta ^*} = \left( {\beta _j^*;\,j = 0,1, \ldots ,n} \right)$ is obtained by solving \eqref{inverVI2-multi}, then our estimator for the cost function is
	\begin{align}
	\hat f\left( x \right) = \sum\limits_{j = 0}^n {\beta _j^*{x^j}}  = 1 + \sum\limits_{j = 1}^n {\beta _j^*{x^j}}.  \label{costEstimator}
	\end{align}
	
	\begin{rem} \label{rem2}
		\em{In the above QP formulations, we have assumed that the parameter vector ${\btheta}$ and the set of user classes $\tilde{\scrU}$ are the same for all $|\scrK|$ networks. We note that in \eqref{costMulti} what essentially gets involved is only the weighed sum of link flows of different types of vehicles (other than the link flow of each single user type). Therefore, as noted in Remark \ref{rem1}, we are very likely to be able to recover the cost functions with satisfactory accuracy from such weighted sum of link flows. In Sec. \ref{sec:num}  we will illustrate this by conducting extensive numerical experiments.}
	\end{rem}

	\begin{algorithm}
		\caption{Method of Successive Averages (MSA) \cite{noriega2007algorithmic}}
		\label{alg:msa}
		\begin{algorithmic}[1]
			\Require the road network $\big( {\tilde\scrV, \tilde\scrA, \tilde\scrW} \big)$; the set of user classes $\tilde{\scrU}$; the function $f(\cdot)$ in \eqref{costMulti}; the demand vectors $\bg^{{(u)}}$, $u \in [\kern-0.15em[ \tilde\scrU 
			]\kern-0.15em]$; a real parameter $\varepsilon > 0$; the maximum iteration times $L$. 
			\State \textbf{Step 0:} Initialization. Initialize link flows $x^{\ell}_{iu} = 0$ for $i \in [\kern-0.15em[ \tilde\scrA]\kern-0.15em], u \in [\kern-0.15em[ \tilde\scrU 
			]\kern-0.15em] $; set iteration counter $\ell=0$.
			\State \textbf{Step 1:} Compute new extremal flows. Set $\ell = \ell + 1$.
			
			\begin{itemize}
				\item[] \textbf{1.1:} Update link travel costs based on current link flows: $t^{\ell}_{iu} = {t_{iu}}( {{x^{\ell-1}_{i1}}, \ldots ,{x^{\ell-1}_{i|\tilde\scrU|}}} ), \, \forall i \in [\kern-0.15em[ \tilde\scrA 
				]\kern-0.15em], u \in [\kern-0.15em[ \tilde\scrU 
				]\kern-0.15em]$
				
				\item[] \textbf{1.2:} Carry out ``all-or-nothing'' assignment of the demands $\bg^{{(u)}}$ on current shortest paths to obtain $y^{\ell}_{iu}$.
				
			\end{itemize}
			\State \textbf{Step 2:} Update link flows via
			\[x^{\ell}_{iu} = x^{\ell-1}_{iu} + {\lambda ^\ell}\left( {y^{\ell}_{iu} - x^{\ell-1}_{iu}} \right),\]
			where ${\lambda ^\ell} = {1 \mathord{\left/
					{\vphantom {1 l}} \right.
					\kern-\nulldelimiterspace} \ell}$.
			
			\State \textbf{Step 3:} Stopping criterion (slightly different than that in \cite{noriega2007algorithmic}).
			Compute the \textit{Relative Gap} ($\textit{RG}$) as
			\[\text{RG} = \frac{{{{\left\| {{\bx^{\ell}} - {\bx^{\ell-1}}} \right\|}}}}{{{{\left\| {{\bx^{\ell}}} \right\|}}}}.\]
			If $\text{RG} < \varepsilon$ or $l \ge L$, terminate; otherwise, return to Step 1.
		\end{algorithmic}
	\end{algorithm}

	\section{Numerical Results}  \label{sec:num}
	
	In all our experimental scenarios, we consider two types of vehicles, cars and trucks, indexed 1 and 2, respectively. Thus, we have $|\tilde\scrU| = 2$ types of vehicles. Taking account of different fraction of contributions to the travel cost, we assume the flow weight vector to be $\btheta = (1.0, 2.0)$; i.e., the truck flows contribute twice of the same amount of car flows to the cost function $f(\cdot)$. In addition, taking into consideration the different degrees of free-flow travel time's dependence on cars and trucks, we adopt ${t^{{0}}_{i1}} = 1.0 \times t^{{0}}_{i}$ and ${t^{{0}}_{i2}} = 1.1 \times t^{{0}}_{i}$, where $t^{{0}}_{i}$ is the \textit{reference free-flow travel time} of the $i$th \emph{physical} link (available via the benchmark datasets that we will use). In what follows, without loss of generality, we only consider $|\scrK| = 1$ for each benchmark network scenario. In particular, for each network, we divide the original demand data proportionally, $80\%$ for cars and $20\%$ for trucks, to obtain the demand matrices for cars and trucks, respectively. The equilibrium link flows are generated by applying the Method of Successive Averages (MSA) \cite{noriega2007algorithmic} (see Alg. \ref{alg:msa}) with parameters $\varepsilon = 10^{-6}$ and $L = 1000$. When recovering the cost function $f(\cdot)$ via invVI-2 (see \eqref{inverVI2-multi}) for each network, we take parameters $n \in \{3, 4, 5, 6\}$, $c \in \{0.5, 1.0, 1.5\}$, and $\gamma \in \{0.01, 0.1, 1.0, 10.0, 100.0\}$. Note that, in practice, the values of $n$, $c$, and $\gamma$ can be determined by cross-validation (see \cite{InverseVIsTraffic}), provided that $|\scrK| > 1$. We also note that the QP invVI-2 can be solved very efficiently even for network incidences with very large sizes. In the following, we consider three benchmark networks, whose sizes range from medium to large.

	\subsection{Sioux-Falls network}
\begin{figure}[h]  
		\centering
		\begin{subfigure}[b]{0.49\textwidth}
			\includegraphics[width=\textwidth]{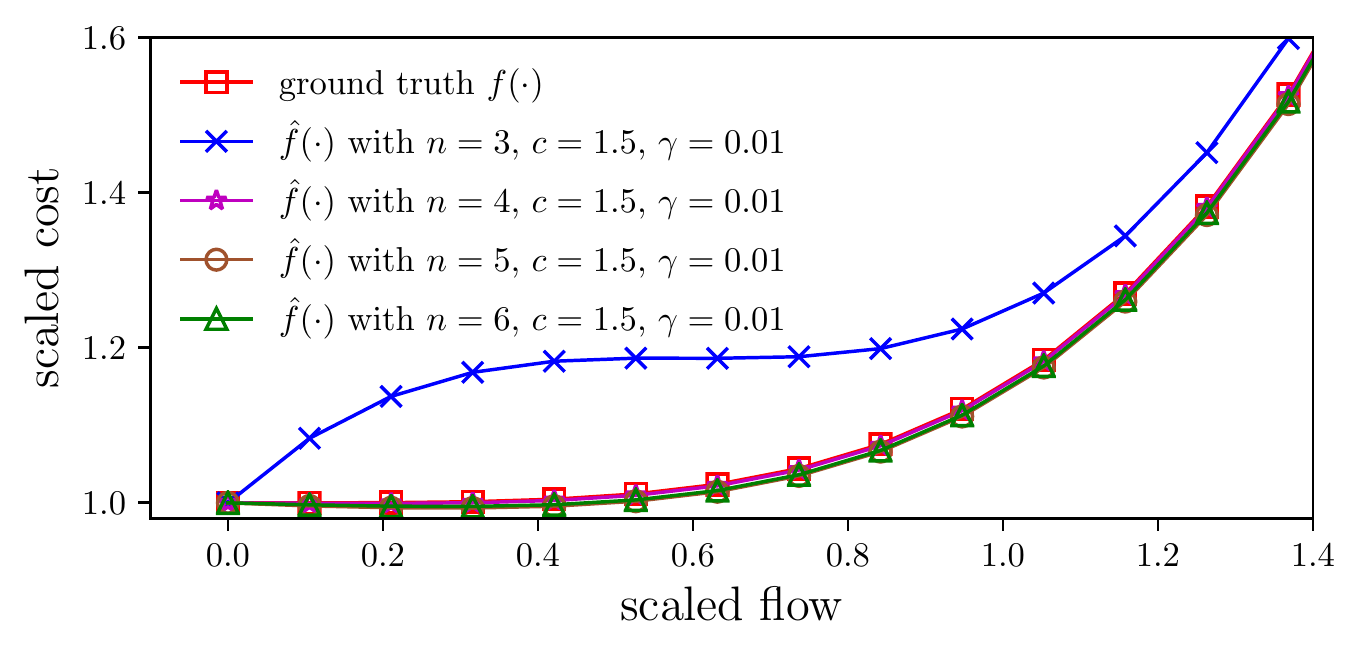}
			\caption{Vary $n$ ($c$ and $\gamma$ fixed)}
			\label{fig:n_Sioux}
		\end{subfigure} 
		\begin{subfigure}[b]{0.49\textwidth}
			\includegraphics[width=\textwidth]{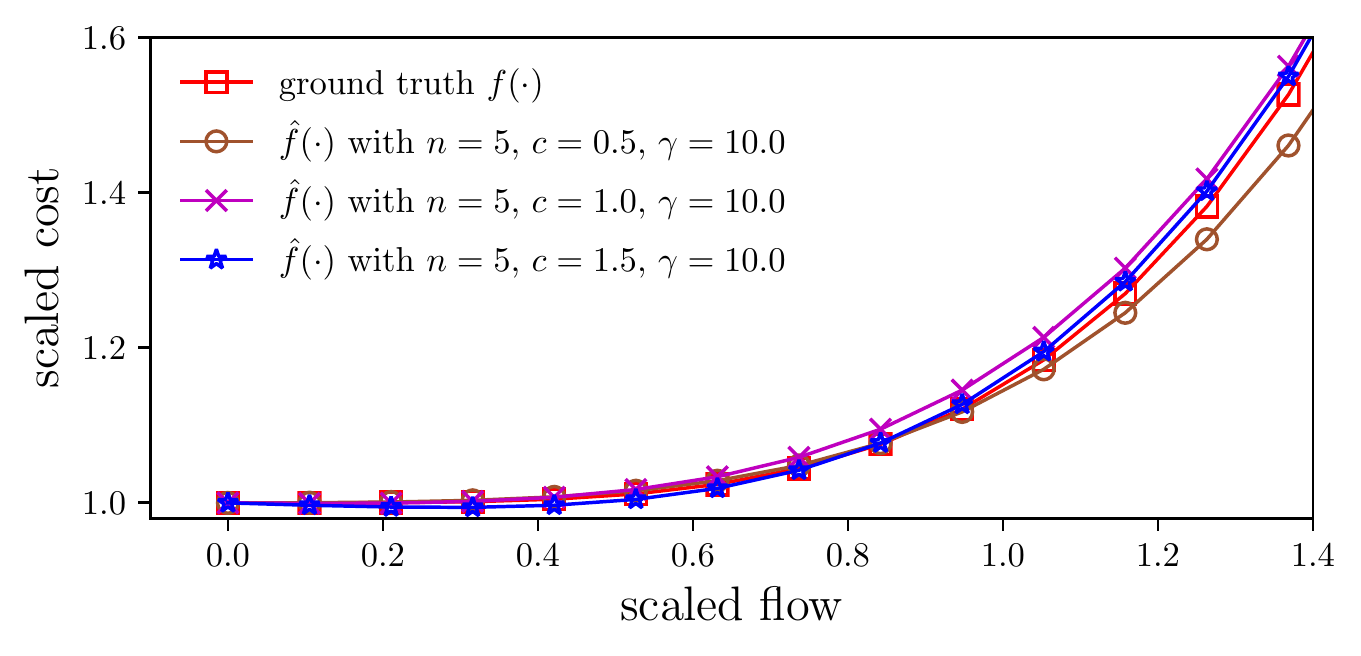}
			\caption{Vary $c$ ($n$ and $\gamma$ fixed)}
			\label{fig:c_Sioux}
		\end{subfigure}   
		\begin{subfigure}[b]{0.49\textwidth}
			\includegraphics[width=\textwidth]{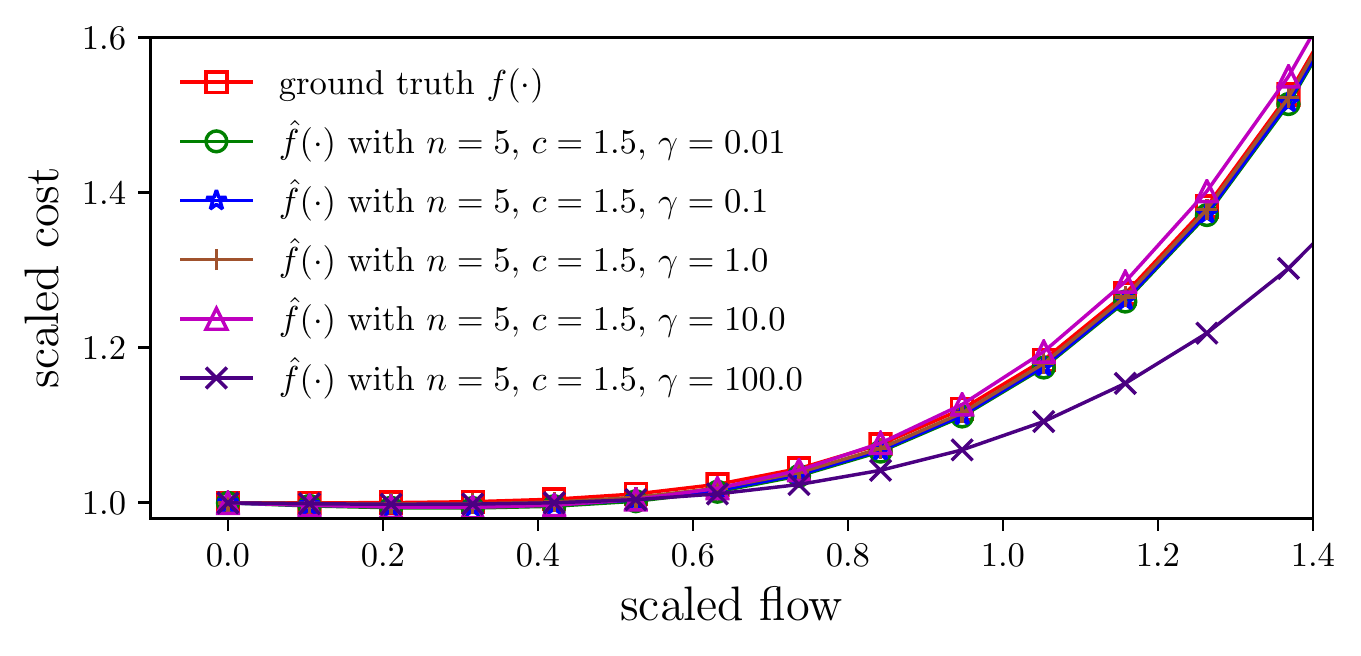}
			\caption{Vary $\gamma$ ($n$ and $c$ fixed)}
			\label{fig:gamma_Sioux}
		\end{subfigure}  
		\caption{Estimations for cost function $f(\cdot)$ by solving invVI-2 corresponding to different parameter settings (Sioux-Falls).}
		\label{fig:Sioux}
	\end{figure}

	The Sioux-Falls network \cite{BarGera16} contains 24 nodes, 24 zones (hence $24 \times (24-1) = 552$ OD pairs), and 76 links. The ground truth $f\left(\cdot\right)$ is taken as $f\left(z\right) = 1 + 0.15z^4, \,z \geq 0$. Fig. \ref{fig:Sioux} shows the estimation results for $f(z)$ by solving invVI-2 corresponding to different parameter settings. In particular, Fig. \ref{fig:n_Sioux} shows the curves of the ground truth $f(z)$ and the estimator $\hat f(z)$ corresponding to $n$ taking values from $\{3, 4, 5, 6\}$ while keeping $c$ and $\gamma$ fixed as 1.5 and 0.01 respectively; it is seen that except for the case $n = 3$, all estimation curves are very close to the ground truth. Note that the ground truth $f(z)$ is a polynomial function with degree 4, which is greater than 3. This suggests that it should be good to use a reasonably bigger $n$ to recover the cost function. Fig. \ref{fig:c_Sioux}
	shows the curves of the ground truth $f(z)$ and the estimator $\hat f(z)$ corresponding to $c$ taking values from $\{0.5, 1.0, 1.5\}$ while keeping $n$ and $\gamma$ fixed as 5 and 10.0 respectively; it is seen that except for the case $c = 0.5$, the estimation curves are very close to the ground truth. This suggests that setting $c$ reasonably bigger should give better estimation results. Fig. \ref{fig:gamma_Sioux}
	plots the curves of the ground truth $f(z)$ and the estimator $\hat f(z)$ corresponding to $\gamma$ taking values from $\{0.01, 0.1, 1.0, 10.0, 100.0\}$ while keeping $n$ and $c$ fixed as 5 and 1.5 respectively; it is seen that except for the case $\gamma = 100.0$, the estimation curves are very close to the ground truth. This suggests that choosing a smaller regularization parameter $\gamma$ should give tighter estimation results in terms of data reconciling.
	
	\begin{figure}[h]  
		\centering
		\begin{subfigure}[b]{0.49\textwidth}
			\includegraphics[width=\textwidth]{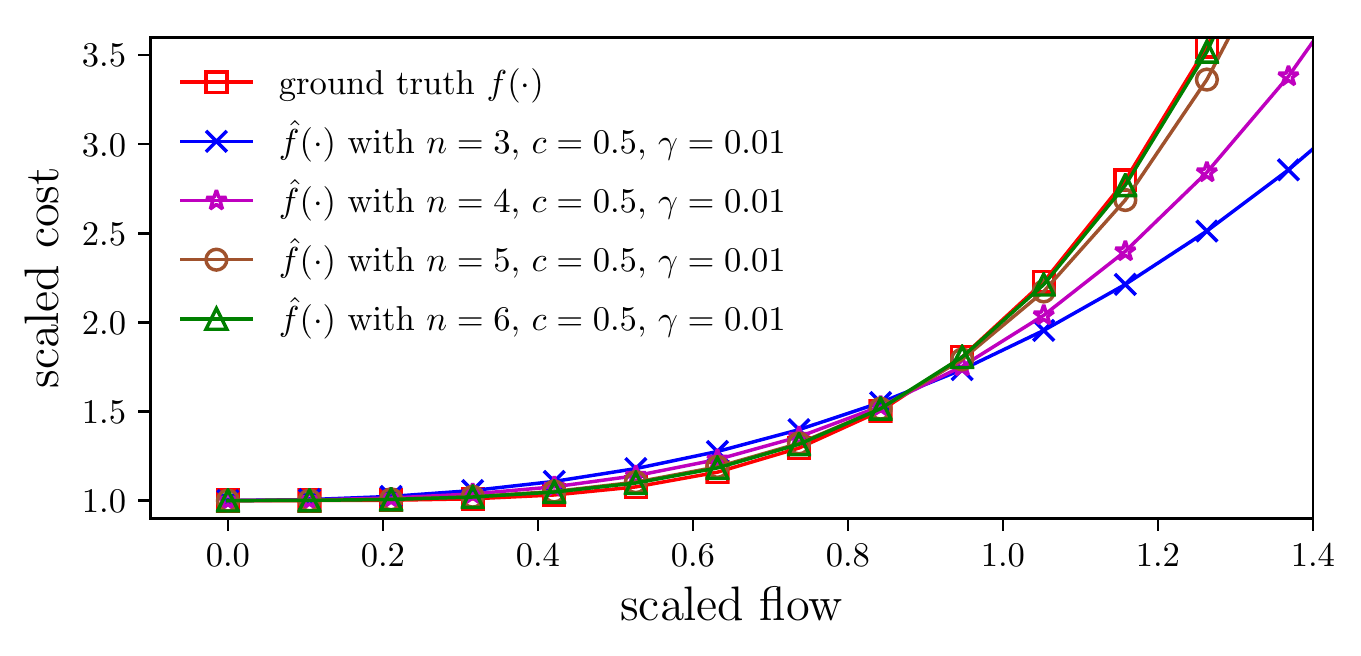}
			\caption{Vary $n$ ($c$ and $\gamma$ fixed)}
			\label{fig:n_Tiergarten}
		\end{subfigure} 
		\begin{subfigure}[b]{0.49\textwidth}
			\includegraphics[width=\textwidth]{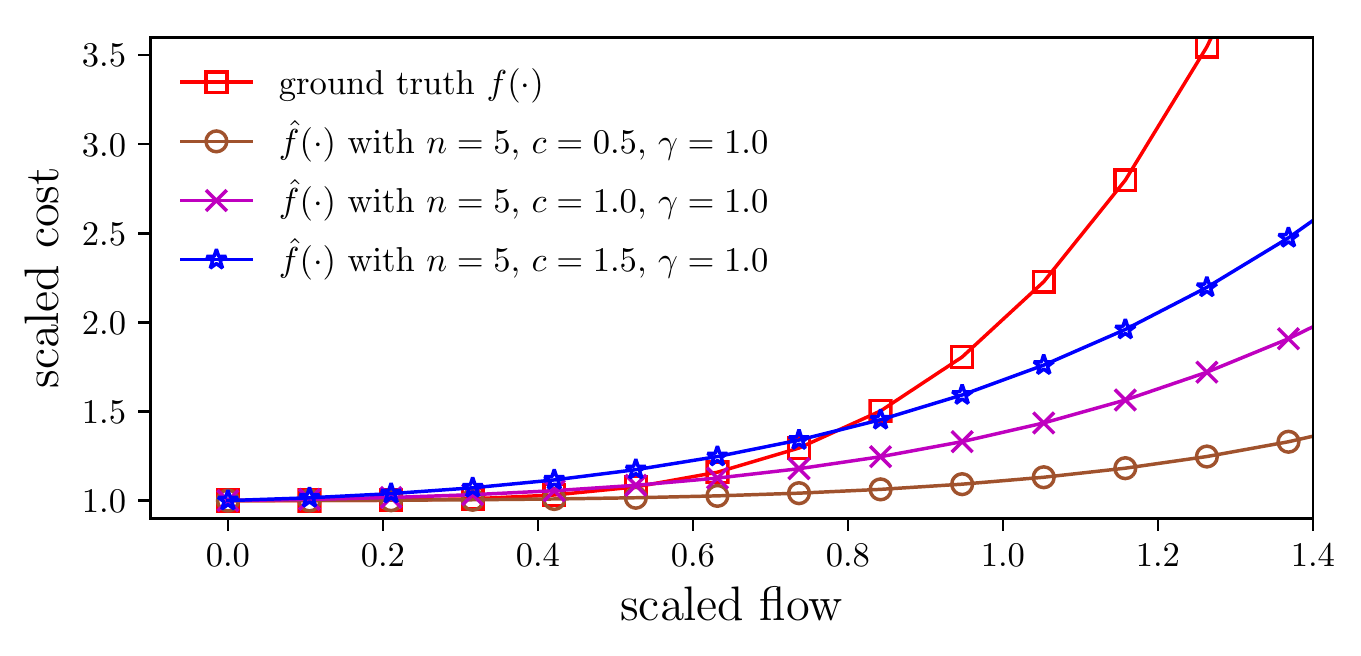}
			\caption{Vary $c$ ($n$ and $\gamma$ fixed)}
			\label{fig:c_Tiergarten}
		\end{subfigure}   
		\begin{subfigure}[b]{0.49\textwidth}
			\includegraphics[width=\textwidth]{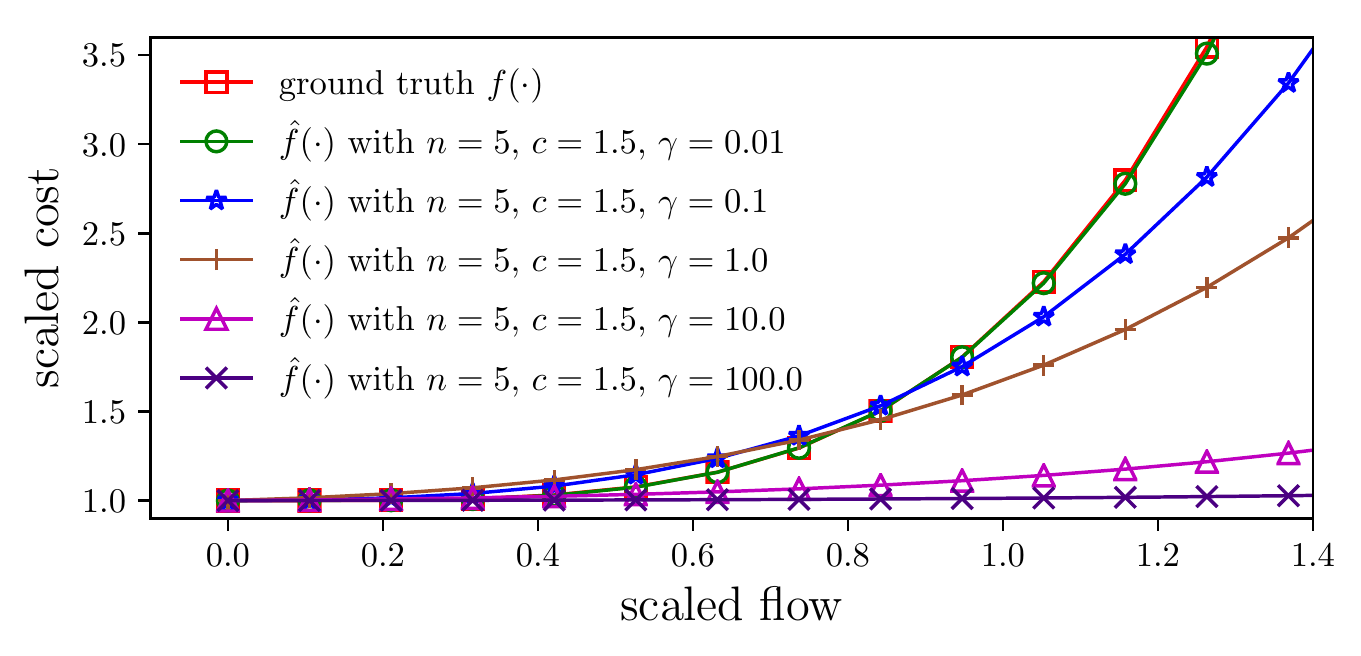}
			\caption{Vary $\gamma$ ($n$ and $c$ fixed)}
			\label{fig:gamma_Tiergarten}
		\end{subfigure}  
		\caption{Estimations for cost function $f(\cdot)$ by solving invVI-2 corresponding to different parameter settings (Tiergarten).}
		\label{fig:Tiergarten}
	\end{figure}

	\begin{figure}[h]  
		\centering
		\begin{subfigure}[b]{0.49\textwidth}
			\includegraphics[width=\textwidth]{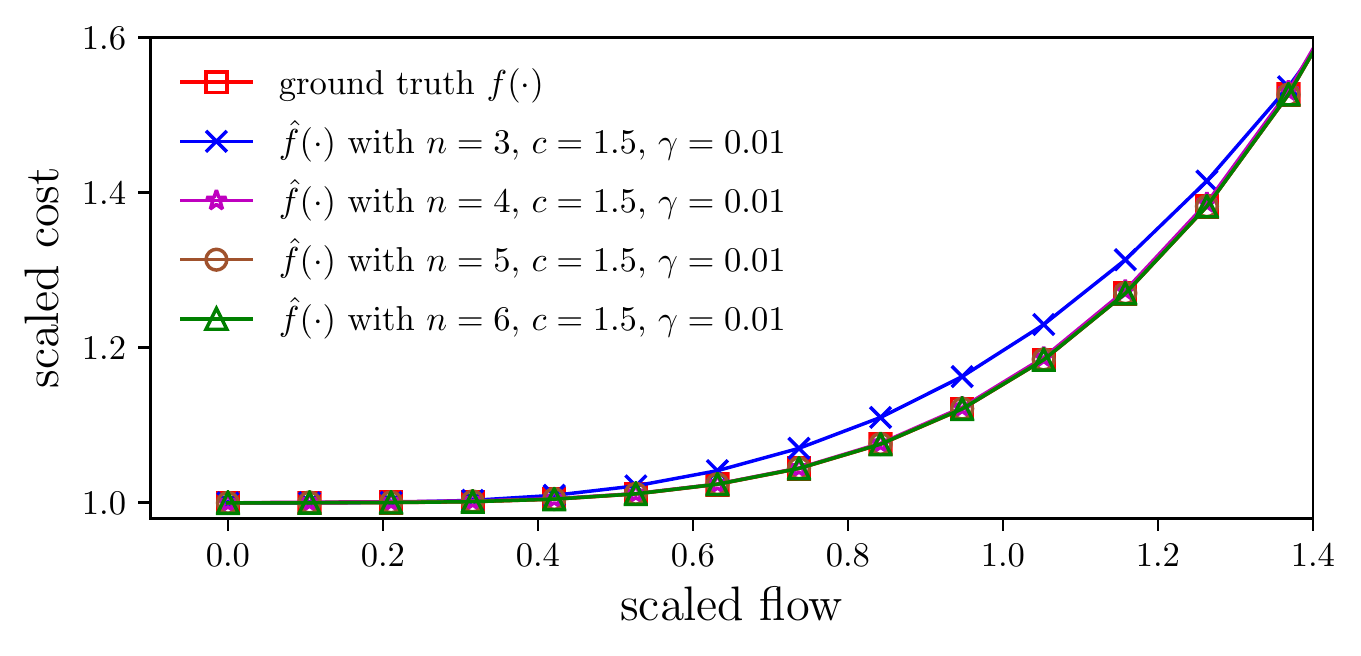}
			\caption{Vary $n$ ($c$ and $\gamma$ fixed)}
			\label{fig:n_Anaheim}
		\end{subfigure} 
		\begin{subfigure}[b]{0.49\textwidth}
			\includegraphics[width=\textwidth]{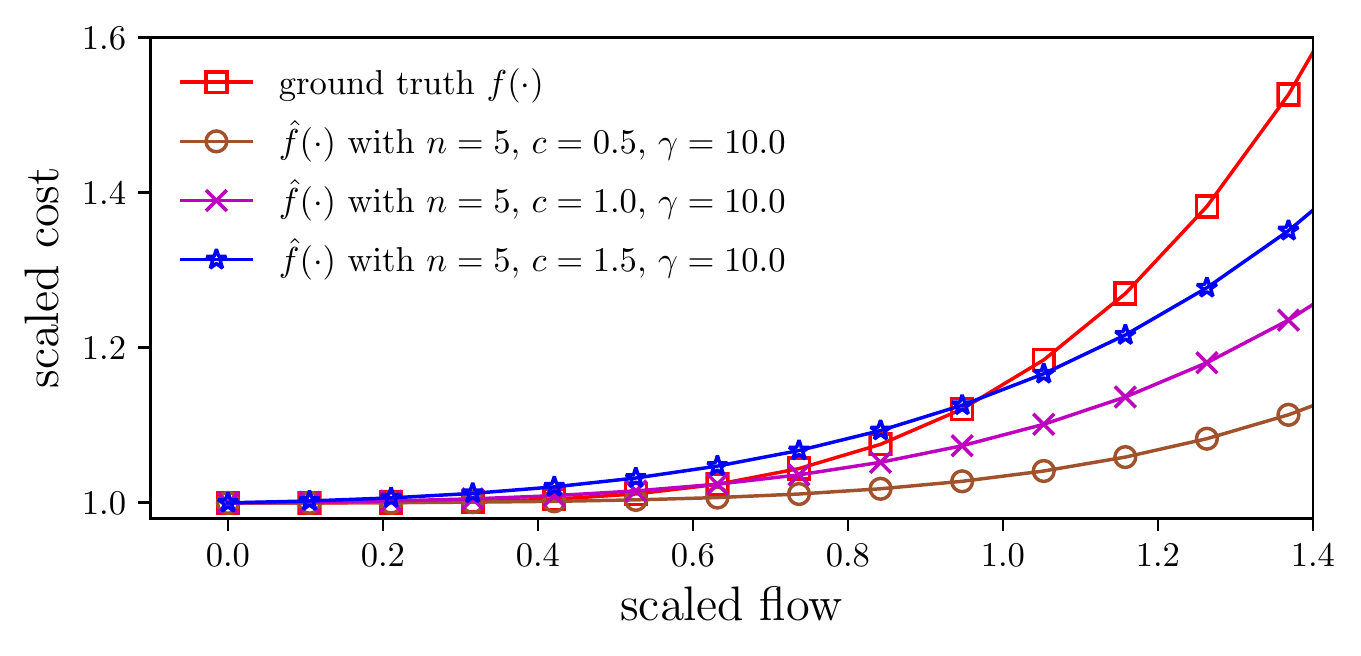}
			\caption{Vary $c$ ($n$ and $\gamma$ fixed)}
			\label{fig:c_Anaheim}
		\end{subfigure}   
		\begin{subfigure}[b]{0.49\textwidth}
			\includegraphics[width=\textwidth]{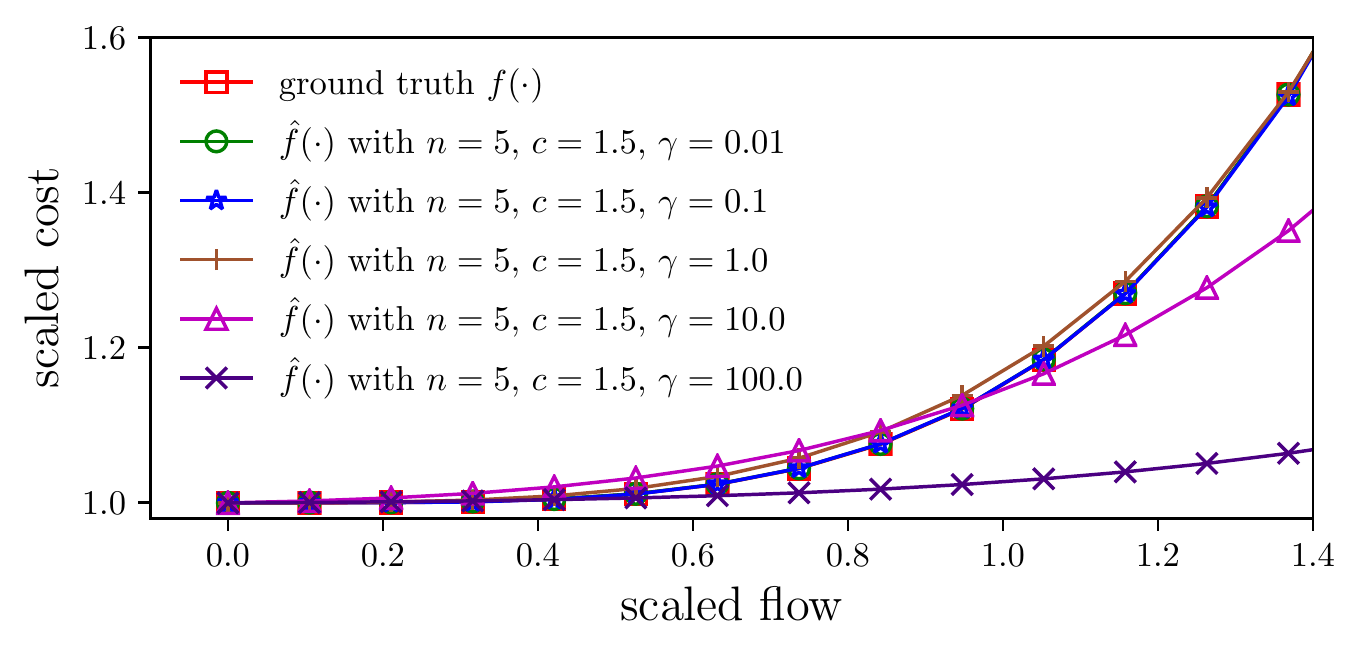}
			\caption{Vary $\gamma$ ($n$ and $c$ fixed)}
			\label{fig:gamma_Anaheim}
		\end{subfigure}  
		\caption{Estimations for cost function $f(\cdot)$ by solving invVI-2 corresponding to different parameter settings (Anaheim).}
		\label{fig:Anaheim}
	\end{figure}

	\subsection{Berlin-Tiergarten network}
	
	The Berlin-Tiergarten network \cite{BarGera16} contains 361 nodes, 26 zones (hence $26 \times (26 - 1) = 650$ OD pairs), and 766 links. The ground truth $f\left(\cdot\right)$ is taken as $f\left(z\right) = 1 + z^4, \,z \geq 0$. Fig. \ref{fig:Tiergarten} shows the curves of the estimator for $f(z)$ by solving invVI-2 corresponding to different parameter settings. In particular, Fig. \ref{fig:n_Tiergarten} shows the curves of the ground truth $f(z)$ and the estimator $\hat f(z)$ corresponding to $n$ taking values from $\{3, 4, 5, 6\}$ while keeping $c$ and $\gamma$ fixed as 0.5 and 0.01 respectively; it is seen that as $n$ increases, the estimation curves get closer and closer to the ground truth. Similar to the Sioux-Falls network, this suggests that it should be good to use a reasonably bigger $n$ to recover the cost function. Fig. \ref{fig:c_Tiergarten}
	shows the curves of the ground truth $f(z)$ and the estimator $\hat f(z)$ corresponding to $c$ taking values from $\{0.5, 1.0, 1.5\}$ while keeping $n$ and $\gamma$ fixed as 5 and 1.0 respectively; it is seen that as $c$ gets bigger and bigger, the estimation curves are closer and closer to the ground truth. Similar to the Sioux-Falls network, this suggests that setting $c$ reasonably bigger should give better estimation results. Fig. \ref{fig:gamma_Tiergarten}
	plots the curves of the ground truth $f(z)$ and the estimator $\hat f(z)$ corresponding to $\gamma$ taking values from $\{0.01, 0.1, 1.0, 10.0, 100.0\}$ while keeping $n$ and $c$ fixed as 5 and 1.5 respectively; it is seen that as $\gamma$ decreases, the estimation curves get closer and closer to the ground truth. Like in the Sioux-Falls network case, this suggests that choosing a smaller regularization parameter $\gamma$ should give tighter estimation results.


	\subsection{Anaheim network}
	
	The Anaheim network \cite{BarGera16} contains 416 nodes, 38 zones (hence $38 \times (38 - 1) = 1406$ OD pairs), and 914 links. The ground truth $f\left(\cdot\right)$ is taken as $f\left(z\right) = 1 + 0.15z^4, \,z \geq 0$. Fig. \ref{fig:Anaheim} plots the graphs of the estimator for $f(z)$ by solving invVI-2 corresponding to different parameter settings. In particular, Fig. \ref{fig:n_Anaheim} shows the curves of the ground truth $f(z)$ and the estimator $\hat f(z)$ corresponding to $n$ taking values from $\{3, 4, 5, 6\}$ while keeping $c$ and $\gamma$ fixed as 1.5 and 0.01 respectively; it is seen that except for the case $n = 3$, all estimation curves are very close to the ground truth. Similar to the Sioux-Falls and the Berlin-Tiergarten networks, this suggests that it should be good to use a reasonably bigger $n$ to recover the cost function. Fig. \ref{fig:c_Anaheim}
	shows the curves of the ground truth $f(z)$ and the estimator $\hat f(z)$ corresponding to $c$ taking values from $\{0.5, 1.0, 1.5\}$ while keeping $n$ and $\gamma$ fixed as 5 and 10.0 respectively; it is seen that, similar to the Berlin-Tiergarten network, as $c$ increases, the estimation curves become closer and closer to the ground truth. This, again, suggests that setting $c$ reasonably bigger should give better estimation results. Fig. \ref{fig:gamma_Anaheim}
	plots the curves of the ground truth $f(z)$ and the estimator $\hat f(z)$ corresponding to $\gamma$ taking values from $\{0.01, 0.1, 1.0, 10.0, 100.0\}$ while keeping $n$ and $c$ fixed as 5 and 1.5 respectively; similar observations can be made as in the case of Berlin-Tiergarten network.

	\section{Conclusions}  \label{sec:con}
	
	In this paper, we investigate the travel latency cost function estimation problem for multi-class transportation networks, where coupled link flows from different types of vehicles are handled. Based on inverse variational inequalities, we propose a generalized data-driven approach (as opposed to the single-class case), whose effectiveness and efficiency are validated by conducting extensive numerical experiments. We show that the approach applies to networks ranging from moderate to larger-sized.



	\bibliographystyle{IEEEtran}
	

	\bibliography{bib1}


\end{document}